\documentclass[prb,twocolumn,groupedaddress,shownopacs,amssymb]{revtex4-2}
\usepackage{graphicx,psfrag,amsmath,amssymb}

\usepackage[usenames, dvipsnames]{color}

\usepackage{hyperref} 
\hypersetup{
    colorlinks=true,
    linkcolor=Blue,
    citecolor=Blue,
    filecolor=Blue,      
    urlcolor=Blue,
    pdftitle={twobody-extended},
    }

\begin{document}

\title{Variational approach for the two-body problem 
\\ in a multiband extended-Hubbard model}

\author{M. Iskin}
\affiliation{
Department of Physics, Ko\c{c} University, Rumelifeneri Yolu, 
34450 Sar\i yer, Istanbul, T\"urkiye
}

\date{\today}

\begin{abstract}

Considering a spin-up and a spin-down fermion in a generic tight-binding 
lattice with a multi-site basis, we investigate the two-body problem using 
a multiband extended-Hubbard model with finite-ranged hopping and 
interaction parameters. We derive a linear eigenvalue problem for the 
entire two-body spectrum, alongside a nonlinear eigenvalue problem for 
the bound states in the form of a self-consistency equation. Our results, 
based on an exact variational approach, suggest potential applications 
across various lattice geometries. As an illustration, we apply them to 
the linear-chain model and show that the resultant spin singlet and 
triplet bound states align well with the existing literature.

\end{abstract}

\maketitle

\section{Introduction}
\label{sec:intro}

Understanding the two-body problem lies at the heart of the BCS 
theory of superconductivity, offering key insights into the microscopic 
mechanisms underlying this phenomenon~\cite{cooper56, bardeen57}. 
For instance, it elucidates how a large number of Cooper pairs 
condense into a single quantum state, leading to the formation 
of an energy gap in the electronic density of states just below 
the Fermi energy and determining the critical temperature for 
pairing~\cite{leggett, leggett80, nozieres85}. 
Moreover, recent investigations have highlighted the crucial role 
of the exactly solvable two-body problem in understanding 
quantum-geometric effects on some other superconducting properties, 
including those of multiband Hubbard lattices, flat-band 
superconductors and spin-orbit coupled Fermi superfluids.
This includes the superfluid weight, superfluid density, velocity 
of the low-energy collective modes, and the kinetic coefficient of 
the Ginzburg-Landau theory but not limited to 
them~\cite{torma18, iskin21, torma22, herzog22, torma23, iskin23, iskin24}. 
Hence, the two-body problem still continues to provide a bottom-up 
approach for untangling the complexities of the many-body 
problem. 
There is no doubt that its further extensions to previously unexplored 
settings may also play fundamental roles~\cite{setty23}, especially
with the emergence of newly discovered superconductors and recent
advances in atomic physics. In the latter context, realizations 
of few-body problems in the cold-atom settings, including both 
the two-body and three-body problems, have stimulated significant
activity~\cite{winkler06, tai17, holten22, naidon17}, and
there is a growing interest in the topological aspects of the 
two-body bound states in various multiband 
lattices~\cite{guo11, gorlach17, marques18, salerno18, lin20, 
zurita20, salerno20, pelegri20, zuo21, okuma23, alyuruk24}. 

In our previous study on generic tight-binding lattices with a 
multi-site basis \cite{iskin21}, the focus was solely on the onsite 
interaction between a spin-up and a spin-down fermion. 
There, we derived a linear eigenvalue problem for the entire 
two-body spectrum and a nonlinear eigenvalue problem for the 
spin-singlet bound states in the form of a self-consistency relation. 
Our expressions were obtained through an exact variational approach 
in reciprocal space, and their application reproduced the results 
found in the literature on the Haldane model which uses exact 
diagonalization in real space~\cite{salerno18}. 
Our self-consistency relation was also derived in subsequent works 
using alternative methods~\cite{orso22, iskin23}. More recently, we 
investigated the evolution of the two-body Hofstadter-Hubbard butterfly 
as a function of interaction strength, and developed an efficient 
formulation for their Chern numbers by utilizing the eigenvectors 
of the nonlinear eigenvalue problem \cite{alyuruk24}. Motivated by 
the success of our previous results on the Hubbard model, here we 
extend the formalism and develop an exact variational approach for 
the two-body problem within the context of a multiband extended-Hubbard 
model with finite-ranged hopping and interaction parameters. 
In contrast to the case of onsite interactions~\cite{iskin21}, 
we discuss the possibility of having both spin-singlet and spin-triplet 
two-body bound states depending on the symmetry and range of the 
interactions.

As an alternative to our variational approach, the 
density-matrix-renormalization-group is a widely used numerical method for 
computing low-lying states in one-dimensional lattices~\cite{schollwock11}. 
It can also be used to obtain the entanglement spectrum, multi-point 
correlators, and real-time dynamics for general one-dimensional systems. 
On the other hand, although the exact-diagonalization method is efficient 
for finding low-lying energies in any dimension~\cite{ed}, it typically 
works with small lattice sizes, which can lead to significant finite-size 
effects. In contrast to them, we would like to emphasize that our 
variational ansatz is designed to be as general as possible, consistent 
with symmetry and conservation principles, and provides exact results 
in the thermodynamic limit, regardless of the signs and magnitudes of 
the model parameters in any dimension.

The remaining sections of this paper are structured as follows. 
In Sec.~\ref{sec:Ham}, we introduce the extended-Hubbard model 
in real space and subsequently transform it into reciprocal space. 
In Sec.~\ref{sec:twobody}, we employ an exact variational approach 
to derive a linear eigenvalue problem for the entire two-body 
spectrum and a nonlinear eigenvalue problem for its bound-state 
branches. In Sec.~\ref{sec:benchmark}, we validate our approach 
by comparing it with the existing literature on the linear-chain 
model. In Sec.~\ref{sec:geometry}, we relate our results to the 
recent literature on quantum-geometric effects.
Finally, in Sec.~\ref{sec:conc}, we provide a summary of our 
findings and offer some outlook for future research.

\section{Lattice Hamiltonian}
\label{sec:Ham}

For spin-$1/2$ fermions with $\sigma = \{\uparrow, \downarrow\}$ denoting
the spin projections, the Hubbard Hamiltonian is typically written as
$
\mathcal{H} = \sum_\sigma \mathcal{H}_\sigma 
+ \mathcal{H}_{\uparrow\downarrow},
$
where $\mathcal{H}_\sigma$ terms describe the corresponding 
single-particle problem for each spin projection and 
$\mathcal{H}_{\uparrow\downarrow}$ term describes the two-body 
interactions between spin-up and spin-down particles~\cite{arovas21, qin21}. 
Within the tight-binding approximation, and considering a generic 
sublattice structure in the lattice, these terms can be written 
in general as
\begin{align}
\mathcal{H}_\sigma &= -\sum_{Si; S'i'} t_{Si; S'i'}^\sigma 
c_{S i \sigma}^\dagger c_{S' i' \sigma},
\\
\mathcal{H}_{\uparrow\downarrow} &= \sum_{Si; S'i'} U_{Si;S'i'}
c_{S i \uparrow}^\dagger c_{S' i' \downarrow}^\dagger 
c_{S' i' \downarrow} c_{S i \uparrow},
\end{align}
where the hopping parameters $t_{Si; S'i'}^\sigma$ describe tunneling of 
a spin-$\sigma$ particle from the sublattice site $S'$ in the unit cell 
$i'$ to the sublattice site $S$ in the unit cell $i$, and the interaction
parameters $U_{Si;S'i'}$ describe the density-density interactions between
a spin-$\uparrow$ particle on site $S \in i$ and a spin-$\downarrow$ particle
on site $S' \in i'$. The range of interactions is assumed to be finite here, 
i.e., we are interested in studying the effects of not only the onsite but 
also the nearest-neighbor, next-nearest-neighbor, etc., interactions on 
the formation of two-body bound states in a generic lattice.
It is worth emphasizing that our variational approach is exact 
for any set of these parameters, regardless of their signs and magnitudes.

Next we reexpress $\mathcal{H}$ in the reciprocal space through the 
canonical transformation~\cite{iskin21}
$
c_{S i \sigma}^\dagger = \frac{1}{\sqrt{N_c}} \sum_\mathbf{k} e^{-i \mathbf{k} \cdot \mathbf{r}_{S i}} c_{S \mathbf{k} \sigma}^\dagger,
$
where $N_c$ is the number of unit cells in the system, 
$\mathbf{k}$ is the crystal momentum (in units of $\hbar \to 1$ the Planck constant) 
in the first Brillouin zone, and $\mathbf{r}_{Si}$ is the position of 
site $S \in i$. This leads to a generic Bloch Hamiltonian of the form
$
\mathcal{H}_\sigma = \sum_{S S' \mathbf{k}} h_{SS'\mathbf{k}}^\sigma
c_{S \mathbf{k} \sigma}^\dagger c_{S' \mathbf{k} \sigma},
$
where the matrix elements $h_{SS'\mathbf{k}}^\sigma$ are defined in the 
sublattice basis through the Fourier transform
$
h_{SS'\mathbf{k}}^\sigma = \frac{1}{N_c}\sum_{ii'} t_{Si;S'i'}^\sigma
e^{\mathrm{i} \mathbf{k} \cdot \mathbf{r}_{Si;S'i'}}
$
with 
$
\mathbf{r}_{Si;S'i'} = \mathbf{r}_{S'i'} - \mathbf{r}_{Si}
$
denoting the relative position. The resultant eigenvalue problem
\begin{align}
\sum_{S'} h_{SS'\mathbf{k}}^\sigma n_{S' \mathbf{k} \sigma} 
= \varepsilon_{n\mathbf{k} \sigma} n_{S \mathbf{k} \sigma}
\end{align}
determines the Bloch bands $\varepsilon_{n\mathbf{k} \sigma}$, 
where $n_{S \mathbf{k} \sigma}$ is the projection of the periodic part
of the corresponding Bloch state onto sublattice $S$. Similarly, 
the interaction term takes the generic form
$
\mathcal{H}_{\uparrow \downarrow} = 
\frac{1}{N_c}\sum_{S S' \mathbf{k} \mathbf{k'} \mathbf{q}} 
U_{SS'}^{\mathbf{k} - \mathbf{k'}}
c_{S, \mathbf{k} + \frac{\mathbf{q}}{2} \uparrow}^\dagger 
c_{S', -\mathbf{k} + \frac{\mathbf{q}}{2}, \downarrow}^\dagger 
c_{S', -\mathbf{k'} + \frac{\mathbf{q}}{2}, \downarrow} 
c_{S, \mathbf{k'} + \frac{\mathbf{q}}{2}, \uparrow},
$
where the amplitudes $U_{SS'}^{\mathbf{k} - \mathbf{k'}}$ of the 
interactions depend on the exchanged momentum $\mathbf{k} - \mathbf{k'}$
through the Fourier transform
$
U_{SS'}^{\mathbf{k} - \mathbf{k'}} = \frac{1}{N_c}\sum_{ii'} U_{Si;S'i'}
e^{\mathrm{i} (\mathbf{k} - \mathbf{k'}) \cdot \mathbf{r}_{Si;S'i'}}.
$
Here we note that
$
U_{SS'}^{\mathbf{k} - \mathbf{k'}} = U_{S'S}^{\mathbf{k'} - \mathbf{k}}
= (U_{SS'}^{\mathbf{k'} - \mathbf{k}})^*
$
must be satisfied by definition. Furthermore, upon transformation 
to the band basis through 
$
c_{S \mathbf{k} \sigma}^\dagger = \sum_n n_{S \mathbf{k} \sigma}^* 
c_{n \mathbf{k} \sigma}^\dagger,
$
the $\mathbf{k}$-space Hamiltonians can be written as~\cite{iskin21}
\begin{align}
\label{eqn:Hsigma}
\mathcal{H}_\sigma &= \sum_{n \mathbf{k}} 
\varepsilon_{n\mathbf{k}\sigma}
c_{n \mathbf{k} \sigma}^\dagger c_{n \mathbf{k} \sigma},
\\
\label{eqn:Hupdown}
\mathcal{H}_{\uparrow\downarrow} &= \frac{1}{N_c} 
\sum_{\substack{nmn'm' \\ \mathbf{k}\mathbf{k'}\mathbf{q}}}
V_{n'm'\mathbf{k'}}^{nm\mathbf{k}}(\mathbf{q})
b_{nm}^\dagger(\mathbf{k}, \mathbf{q})
b_{n'm'}(\mathbf{k'}, \mathbf{q}),
\end{align}
where the amplitudes $V_{n'm'\mathbf{k'}}^{nm\mathbf{k}}(\mathbf{q})$
of the interactions are given in general by
$
V_{n'm'\mathbf{k'}}^{nm\mathbf{k}}(\mathbf{q}) = \sum_{SS'} 
U_{SS'}^{\mathbf{k} - \mathbf{k'}}
n_{S, \mathbf{k}+\frac{\mathbf{q}}{2}, \uparrow}^*
m_{S', -\mathbf{k}+\frac{\mathbf{q}}{2}, \downarrow}^*
{m'}_{S', -\mathbf{k'}+\frac{\mathbf{q}}{2}, \downarrow}
{n'}_{S, \mathbf{k'}+\frac{\mathbf{q}}{2}, \uparrow},
$
and the operator
$
b_{nm}^\dagger (\mathbf{k}, \mathbf{q}) = 
c_{n,\mathbf{k}+\frac{\mathbf{q}}{2}, \uparrow}^\dagger
c_{m,-\mathbf{k}+\frac{\mathbf{q}}{2}, \downarrow}^\dagger
$
creates a pair of $\uparrow$ and $\downarrow$ particles in the 
corresponding Bloch bands with a relative momentum $\mathbf{k}$ and 
a total momentum $\mathbf{q}$.

\section{Two-body problem}
\label{sec:twobody}

Having in mind a multiband lattice Hamiltonian that is invariant under 
discrete translations, the exact solutions for the two-body problem, 
i.e., for any given center-of-mass momentum $\mathbf{q}$, 
can in general be obtained through the variational ansatz
$
|\Psi(\mathbf{q}) \rangle = \sum_{n m \mathbf{k} \sigma \sigma'} 
\alpha_{nm\mathbf{k}}^{\sigma \sigma'}(\mathbf{q}) 
c_{n, \mathbf{k}+\frac{\mathbf{q}}{2}, \sigma}^\dagger 
c_{m,-\mathbf{k}+\frac{\mathbf{q}}{2},\sigma'}^\dagger 
| 0 \rangle, 
$
where $|0 \rangle$ represents the particle vacuum. Here the variational 
parameters must satisfy
$
\alpha_{nm\mathbf{k}}^{\sigma \sigma'} (\mathbf{q}) 
= - \alpha_{mn,-\mathbf{k}}^{\sigma' \sigma} (\mathbf{q}) 
$
so that $|\Psi(\mathbf{q}) \rangle$ is anti-symmetric under fermion 
exchange. Furthermore, given the absence of a spin-orbit-coupling 
term in the single-particle Hamiltonian, they must satisfy 
$
\alpha_{nm\mathbf{k}}^{\sigma \sigma'} (\mathbf{q}) 
= \pm \alpha_{mn,-\mathbf{k}}^{\sigma \sigma'} (\mathbf{q}) 
= \mp \alpha_{nm\mathbf{k}}^{\sigma' \sigma} (\mathbf{q})
$
for the spin-singlet and spin-triplet states, respectively. These 
conditions guarantee that the singlet states are symmetric 
(anti-symmetric) but the triplet states are anti-symmetric (symmetric) 
under spatial (spin) exchange. For the simplicity of presentation, 
here we choose~\cite{iskin21}
\begin{align}
|\psi_\mathbf{q} \rangle = \sum_{n m \mathbf{k}} 
\alpha_{nm\mathbf{k}}^\mathbf{q} 
c_{n, \mathbf{k}+\frac{\mathbf{q}}{2}, \uparrow}^\dagger 
c_{m,-\mathbf{k}+\frac{\mathbf{q}}{2},\downarrow}^\dagger 
| 0 \rangle,
\label{eqn:ansatz}
\end{align}
where 
$
\alpha_{nm\mathbf{k}}^\mathbf{q} \equiv 
\alpha_{nm\mathbf{k}}^{\uparrow \downarrow} (\mathbf{q}) 
$
parameters satisfy
$
\alpha_{nm\mathbf{k}}^\mathbf{q} = \pm \alpha_{mn,-\mathbf{k}}^\mathbf{q}
$
for the singlet and triplet states, respectively. 
They are in such a way that 
$
|\psi_\mathbf{q} \rangle \to \mp |\psi_\mathbf{q} \rangle  
$
upon the transformation $\uparrow \leftrightarrow \downarrow$, 
corresponding, respectively, to an anti-symmetric and symmetric 
combination, i.e., 
$
\frac{|\uparrow\downarrow\rangle \mp |\downarrow\uparrow \rangle}{\sqrt{2}},
$
for the singlet and triplet states under spin exchange.

For any given $\mathbf{q}$, the exact two-body energies $E_\mathbf{q}$ 
are determined by minimizing the expectation value 
$
\langle \psi_\mathbf{q} | \mathcal{H} - E_\mathbf{q} |\psi_\mathbf{q} \rangle
$
with respect to $\alpha_{nm\mathbf{k}}^\mathbf{q}$~\cite{iskin21}. 
This leads to a set of linear equations
\begin{align}
\big(\varepsilon_{n, \mathbf{k} +\frac{\mathbf{q}}{2}, \uparrow}
&+ \varepsilon_{m, -\mathbf{k}+\frac{\mathbf{q}}{2},\downarrow} 
- E_\mathbf{q} \big)
\alpha_{n m \mathbf{k}}^\mathbf{q} \nonumber \\
& +\frac{1}{N_c} \sum_{n' m' \mathbf{k'}}
V_{n'm'\mathbf{k'}}^{nm\mathbf{k}}(\mathbf{q}) 
\alpha_{n' m' \mathbf{k'}}^\mathbf{q} = 0,
\label{eqn:alphanmk}
\end{align}
from which $E_\mathbf{q}$ can be determined as the eigenvalues of an 
$N_b^2 N_c \times N_b^2 N_c$ matrix, where $N_b$ is the number of 
sublattice sites in a unit cell, i.e., the total number of lattice 
sites in the system is $N_b N_c$. Note that 
$
\alpha_{nm \mathbf{k}}^\mathbf{q} \to \pm \alpha_{nm, -\mathbf{k}}^\mathbf{q} 
$
upon spin exchange when $\uparrow \leftrightarrow \downarrow$.
Since the solutions of Eq.~(\ref{eqn:alphanmk}) give the entire 
two-body spectrum, it does not discriminate between the scattering 
(i.e., continuum) and the bound states. As an alternative description, 
we define a set of dressed parameters
\begin{align}
\beta_{SS' \mathbf{k}}^\mathbf{q} 
= \sum_{n m \mathbf{k'}}
U_{SS'}^{\mathbf{k} - \mathbf{k'}}
{n}_{S, \mathbf{k'}+\frac{\mathbf{q}}{2}, \uparrow}
{m}_{S', -\mathbf{k'}+\frac{\mathbf{q}}{2}, \downarrow}
\alpha_{n m \mathbf{k'}}^\mathbf{q},
\end{align}
which are in such a way that
$
\beta_{SS' \mathbf{k}}^\mathbf{q} \to \pm \beta_{S'S, -\mathbf{k}}^\mathbf{q} 
$
upon spin exchange when $\uparrow \leftrightarrow \downarrow$. 
It turns out these dressed parameters are non-zero only for the 
two-body bound states, i.e., they play the role of an order parameter 
for pairing. See the related discussion at the end of this section. 
In more general terms, one may define
$
\beta_{SS' \mathbf{k}}^{\sigma \sigma'} (\mathbf{q}) 
= \sum_{n m \mathbf{k'}}
U_{SS'}^{\mathbf{k} - \mathbf{k'}}
{n}_{S, \mathbf{k'}+\frac{\mathbf{q}}{2}, \sigma}
{m}_{S', -\mathbf{k'}+\frac{\mathbf{q}}{2}, \sigma'}
\alpha_{n m \mathbf{k'}}^{\sigma \sigma'} (\mathbf{q}),
$
where
$
\beta_{SS' \mathbf{k}}^\mathbf{q} 
\equiv \beta_{SS' \mathbf{k}}^{\uparrow \downarrow} (\mathbf{q})
$
is our dressed parameter. Given that they must satisfy
$
\beta_{SS' \mathbf{k}}^{\downarrow \uparrow} (\mathbf{q}) 
= - \beta_{S'S, -\mathbf{k}}^{\uparrow \downarrow} (\mathbf{q})
$
under fermion exchange, we require 
$
\beta_{SS' \mathbf{k}}^\mathbf{q} = \pm \beta_{S'S, -\mathbf{k}}^\mathbf{q}
$
for the singlet and triplet states, respectively. Note that, in the 
presence of onsite interactions only~\cite{iskin21}, 
i.e., when the interaction amplitudes
$
U_{SS'}^{\mathbf{k} - \mathbf{k'}} = U_S \delta_{SS'}
$
are constants in $\mathbf{k}$ space for the intra-orbital 
interactions and vanish for the inter-orbital ones, only the singlet 
bound states are allowed since the order parameter 
for the triplet pairs 
$
\beta_{SS' \mathbf{k}}^\mathbf{q} \to \beta_S^\mathbf{q} 
= -\beta_S^\mathbf{q} 
$
must vanish by the symmetry requirement. Here $\delta_{ij}$ is a 
Kronecker delta. With these definitions, Eq.~(\ref{eqn:alphanmk}) 
reduces to a set of coupled integral equations
\begin{align}
\beta_{\bar{S}\bar{S}' \mathbf{k}}^\mathbf{q} 
= -\frac{1}{N_c} \sum_{n m \mathbf{k'} S S'}
& \frac{U_{\bar{S} \bar{S}'}^{\mathbf{k} - \mathbf{k'}}
{m}_{\bar{S}', -\mathbf{k'}+\frac{\mathbf{q}}{2}, \downarrow}
{n}_{\bar{S}, \mathbf{k'}+\frac{\mathbf{q}}{2}, \uparrow}
}
{\varepsilon_{n, \mathbf{k'} +\frac{\mathbf{q}}{2}, \uparrow}
+ \varepsilon_{m, -\mathbf{k'}+\frac{\mathbf{q}}{2},\downarrow} 
- E_{e \mathbf{q}}}
\nonumber \\
\times & n_{S, \mathbf{k'}+\frac{\mathbf{q}}{2}, \uparrow}^*
m_{S', -\mathbf{k'}+\frac{\mathbf{q}}{2}, \downarrow}^*
\beta_{SS' \mathbf{k'}}^\mathbf{q},
\label{eqn:betaSS}
\end{align}
from which the bound-state energies $E_{e \mathbf{q}}$ can be 
determined through heavy numerics. Note that Eq.~(\ref{eqn:betaSS}) 
reduces to a self-consistency relation when
$U_{SS'}^{\mathbf{k} - \mathbf{k'}}$ is independent of momentum, 
i.e., in the case of usual Hubbard model with onsite 
interactions~\cite{iskin21}.

In order to simplify Eq.~(\ref{eqn:betaSS}) and make further analytical 
progress, next we express $U_{SS'}^{\mathbf{k} - \mathbf{k'}}$ as a
linear combination of different pairing channels, i.e., separable 
functions of $\mathbf{k}$ and $\mathbf{k'}$ in the form
\begin{align}
U_{SS'}^{\mathbf{k} - \mathbf{k'}} = \sum_\ell C_{SS'}^\ell
[\Gamma_{SS'}^\ell(\mathbf{k})]^* \Gamma_{SS'}^\ell(\mathbf{k'}),
\end{align}
where the momentum-independent coefficients $C_{SS'}^\ell$ are 
determined by the interaction parameters $U_{Si;S'i'}$.
For a given $SS'$ sector, it proves convenient to choose the symmetry 
functions $\Gamma_{SS'}^\ell(\mathbf{k})$ in such a way that they satisfy
$
\sum_\mathbf{k} [\Gamma_{SS'}^\ell(\mathbf{k})]^* 
\Gamma_{SS'}^{\ell'}(\mathbf{k}) = \kappa_{SS'}^\ell \delta_{\ell \ell'},
$
i.e., the pairing channels are linearly independent from each other. 
Note that the Hermiticity requirement
$
\mathcal{H}_{\uparrow \downarrow} = 
\mathcal{H}_{\uparrow \downarrow}^\dagger
$
for the Hamiltonian under adjoint operation leads to
$
V_{n'm'\mathbf{k'}}^{nm\mathbf{k}}(\mathbf{q}) 
= [V_{nm\mathbf{k}}^{n'm'\mathbf{k'}}(\mathbf{q})]^*,
$
suggesting that
$
C_{SS'}^\ell = (C_{SS'}^\ell)^*
$
is a real parameter. In addition, the invariance requirement
$
\mathcal{H}_{\uparrow \downarrow} = 
\mathcal{H}_{\downarrow \uparrow}
$
for the Hamiltonian under spin exchange leads to
$
V_{n'm'\mathbf{k'}}^{nm\mathbf{k}}(\mathbf{q})
= V_{m'n',-\mathbf{k'}}^{mn,-\mathbf{k}}(\mathbf{q}),
$
suggesting that 
$
U_{SS'}^{\mathbf{k} - \mathbf{k'}} = U_{S'S}^{\mathbf{k'} - \mathbf{k}}.
$
Given that $C_{SS'}^\ell = C_{S'S}^\ell$ parameters can always be 
chosen symmetrically under sublattice exchange, the latter condition 
allows two distinct solutions
$
\Gamma_{SS'}^\ell(\mathbf{k}) = \pm \Gamma_{S'S}^\ell(-\mathbf{k}),
$
leading to
$
\kappa_{SS'}^\ell = \kappa_{S'S}^\ell
$
as well. In terms of these symmetry functions, the dressed parameters 
can be reexpressed in general as
\begin{align}
\beta_{SS' \mathbf{k}}^\mathbf{q} = \sum_\ell \Lambda_{S S'}^{\ell \mathbf{q}}
[\Gamma_{SS'}^\ell(\mathbf{k})]^*,
\label{eqn:beta}
\end{align}
where the $\mathbf{k}$-independent prefactor can be written as
$
\Lambda_{S S'}^{\ell \mathbf{q}} = C_{SS'}^\ell \sum_{nm\mathbf{k}}
\Gamma_{SS'}^{\ell nm}(\mathbf{k}, \mathbf{q}) 
\alpha_{n m \mathbf{k}}^\mathbf{q}
$
with 
$
\Gamma_{SS'}^{\ell nm}(\mathbf{k}, \mathbf{q})
= \Gamma_{SS'}^\ell(\mathbf{k})
{n}_{S, \mathbf{k}+\frac{\mathbf{q}}{2}, \uparrow}
{m}_{S', -\mathbf{k}+\frac{\mathbf{q}}{2}, \downarrow}.
$
Thus, for any given pairing channel $\ell$, Eq.~(\ref{eqn:beta}) 
suggests that the singlet and triplet states are characterized by
$
\Gamma_{SS'}^\ell(\mathbf{k}) = \pm \Gamma_{S'S}^\ell(-\mathbf{k}),
$
respectively, and 
$
\Lambda_{S S'}^{\ell \mathbf{q}} = \Lambda_{S' S}^{\ell \mathbf{q}}
$
is symmetric under sublattice exchange. Furthermore, the requirement 
$
U_{SS'}^{\mathbf{k} - \mathbf{k'}} = (U_{S'S}^{\mathbf{k} - \mathbf{k'}})^*
$
suggests that 
$
\Gamma_{SS'}^\ell(\mathbf{k}) = \pm [\Gamma_{SS'}^\ell(-\mathbf{k})]^*
$
for the singlet and triplet states, respectively.
By plugging Eq.~(\ref{eqn:beta}) into Eq.~(\ref{eqn:betaSS}), we find 
a set of nonlinear equations in the form of a self-consistency relation
\begin{align}
\Lambda_{\bar{S} \bar{S}'}^{\ell \mathbf{q}} = 
-\frac{C_{\bar{S} \bar{S}'}^\ell}{N_c} 
\sum_{\substack{nm\mathbf{k} \\ SS'\ell'}}
\frac{\Gamma_{\bar{S} \bar{S}'}^{\ell nm}(\mathbf{k}, \mathbf{q})
[\Gamma_{SS'}^{\ell' nm}(\mathbf{k}, \mathbf{q})]^* 
}
{\varepsilon_{n, \mathbf{k} +\frac{\mathbf{q}}{2}, \uparrow}
+ \varepsilon_{m, -\mathbf{k}+\frac{\mathbf{q}}{2},\downarrow} 
- E_{e \mathbf{q}}} 
\Lambda_{S S'}^{\ell' \mathbf{q}},
\label{eqn:LambdaSS}
\end{align}
from which the bound-state energies $E_{e \mathbf{q}}$ can be 
determined efficiently through low-cost numerics.

We note in passing that a suggestive way of expressing the interaction 
amplitude $V_{n'm'\mathbf{k'}}^{nm\mathbf{k}}(\mathbf{q})$ in the 
band basis is
$
V_{n'm'\mathbf{k'}}^{nm\mathbf{k}}(\mathbf{q}) = \sum_{SS' \ell}
C_{SS'}^\ell
[\Gamma_{SS'}^{\ell nm}(\mathbf{k}, \mathbf{q})]^*
\Gamma_{SS'}^{\ell n'm'}(\mathbf{k'}, \mathbf{q}).
$
Then, Eq.~(\ref{eqn:LambdaSS}) resembles the self-consistency equation 
that appears in the BCS theory of superconductivity. We also note that 
a suggestive way of expressing the dressed parameters is
$
\beta_{SS' \mathbf{k}}^{\uparrow \downarrow} (\mathbf{q}) 
= \sum_\mathbf{k'} U_{SS'}^{\mathbf{k} - \mathbf{k'}}
\langle 0 | 
c_{S, \mathbf{k'} + \frac{\mathbf{q}}{2}, \uparrow}
c_{S', -\mathbf{k'} + \frac{\mathbf{q}}{2}, \downarrow} 
| \psi_\mathbf{q} \rangle
= - \beta_{S'S, -\mathbf{k}}^{\downarrow \uparrow} (\mathbf{q}),
$
where $| \psi_\mathbf{q} \rangle$ is the two-body ansatz given
in Eq.~(\ref{eqn:ansatz}). In comparison, considering stationary 
Cooper pairs with $\mathbf{q} = \mathbf{0}$, the BCS order parameters 
for the multi-sublattice Hamiltonian can be written as
$
\Delta_{SS' \mathbf{k}}^{\uparrow \downarrow} (\mathbf{0})
= \sum_\mathbf{k'} U_{SS'}^{\mathbf{k} - \mathbf{k'}}
\langle \psi_\mathrm{BCS} | 
c_{S \mathbf{k'} \uparrow}
c_{S', -\mathbf{k'}, \downarrow} 
| \psi_\mathrm{BCS} \rangle
= - \Delta_{S'S, -\mathbf{k}}^{\downarrow \uparrow} (\mathbf{0}),
$
where $| \psi_\mathrm{BCS} \rangle$ is the coherent BCS ground 
state~\cite{Tsuneto_1998}. Thus, the number conserving expectation value
$
\langle 0 |\cdots| \psi_\mathbf{q} \rangle
$
plays precisely the role of the so-called anomalous average 
$
\langle \psi_\mathrm{BCS} | \cdots | \psi_\mathrm{BCS} \rangle
$
in the BCS theory. In other words, our variational parameters
$
\alpha_{n m \mathbf{k}}^\mathbf{q}
$ 
reduce to the Leggett's number-conserving variational BCS parameter 
$
F_\mathbf{k} \equiv \alpha_{\mathbf{k}}^\mathbf{0}
$
in the case of a single-band continuum system~\cite{leggett}.

\section{Numerical Benchmark}
\label{sec:benchmark}

\begin{figure} [htb]
\includegraphics[width = 0.5\linewidth]{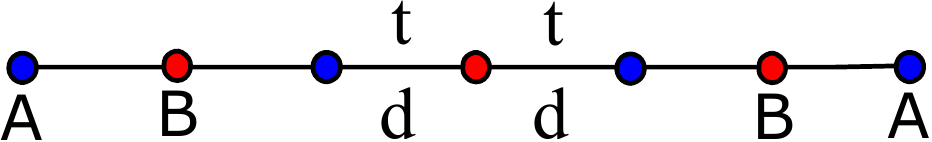}
\caption{\label{fig:chain}
Simulation of the usual linear chain as a lattice with a two-point basis, 
where $S = (A, B)$ denotes the underlying sublattices, $d$ is the 
lattice spacing and $t > 0$ is the nearest-neighbor hopping parameter. 
Note that the reduced first BZ 
$
-\frac{\pi}{2d} \le k_x < \frac{\pi}{2d}
$
is folded into two in comparison to that of the usual linear chain.
}
\end{figure}

To benchmark our approach with the existing 
literature~\cite{nguenang09, valiente09, kornilovitch24}, 
next we simulate the well-studied usual linear chain as a lattice with 
a two-point basis, i.e., with $N_b = 2$. 
This model is illustrated in Fig.~\ref{fig:chain}, where the nearest-neighbor 
hopping parameter is taken as $t > 0$ uniformly across the lattice for 
both spin-up and spin-down particles, i.e., the lattice sites belonging 
to sublattices $A$ and $B$ are identical. Assuming periodic boundary 
conditions, the Bloch Hamiltonian is governed simply by the matrix 
elements
$
h_{AB \mathbf{k}}^\sigma = h_{BA \mathbf{k}}^\sigma = -2t \cos(k_x d)
$
and
$
h_{AA \mathbf{k}}^\sigma = h_{BB \mathbf{k}}^\sigma = 0,
$
and the reduced first Brillouin zone (BZ) is given by
$
-\frac{\pi}{2d} \le k_x < \frac{\pi}{2d},
$
where $d$ is the lattice spacing. Since there are precisely $N_c$ 
states in the BZ, the length $L$ of the simulated lattice is in such 
a way that $L/d = N_b N_c$ gives the total number of sites. 
Thus, a compact way to express the upper ($s = +$) and lower ($s = -$) 
Bloch bands is
$
\varepsilon_{s \mathbf{k} \sigma} = s 2t \cos(k_x d),
$
where the projections
$
s_{A \mathbf{k} \sigma} = 1/\sqrt{2} 
$
and
$
s_{B \mathbf{k} \sigma} = -s/\sqrt{2}
$
determine the associated Bloch states. 

\begin{figure} [htb]
\includegraphics[width = 0.99\linewidth]{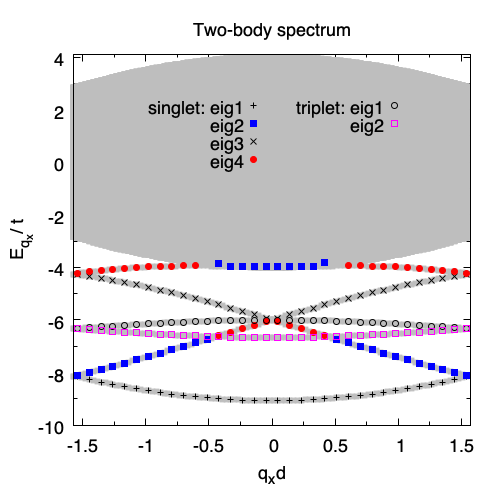}
\caption{\label{fig:2body}
Two-body spectrum $E_{q_x}$ for the linear chain in the reduced BZ.
Here $U = V = -6t$ for the onsite and nearest-neighbor interactions, 
respectively. Full spectrum follows from Eq.~(\ref{eqn:alphanmk})
with $N_c = 101$, and it is shown in gray. Singlet and triplet 
bound-state branches follow from Eq.~(\ref{eqn:LambdaSS}) where 
eig$e$ refers to $E_{e \mathbf{q}}$. 
Note that the entire spectrum appears as folded into the BZ, 
e.g., there appears $4$ $(2)$ instead of $2$ $(1)$ singlet (triplet) 
branches.
}
\end{figure}

Similar to the existing literature, here we consider only the onsite 
($U$) and nearest-neighbor ($V$) interactions, leading to
$
U_{AA}^{\mathbf{k} - \mathbf{k'}} = U = U_{BB}^{\mathbf{k} - \mathbf{k'}}
$
contribution for the intra-sublattice interactions and
$
U_{AB}^{\mathbf{k} - \mathbf{k'}} = 2V\cos(k_xd-k_x'd)
= U_{BA}^{\mathbf{k} - \mathbf{k'}}
$
for the inter-sublattice ones. The two-body spectrum that is shown in 
gray color in Fig.~\ref{fig:2body} is obtained by plugging these 
expressions into Eq.~(\ref{eqn:alphanmk}) with $U = V = -6t$, 
corresponding to attractive interactions. It is important to remark 
that, by construction, our approach produces exact results for any 
signs or strengths of $U$ and $V$. 
Furthermore, in order to minimize possible finite-size effects, 
we chose a very long chain with $N_c = 101$ repeating unit cells. 
We also verified numerically that increasing $N_c$ does not produce 
any distinguishable effect on the presented results. Thus, our results 
are numerically exact for a thermodynamic system.

In addition to a broad region of continuum states, Fig.~\ref{fig:2body}
shows six two-body bound-state branches in the folded BZ. 
To distinguish spin singlet branches from the triplet 
ones, next we construct the appropriate symmetry functions and employ 
them in Eq.~(\ref{eqn:LambdaSS}). In accordance with the analysis given 
in Sec.~\ref{sec:twobody}, 
$
\Gamma_{SS}^\ell(\mathbf{k}) = \pm \Gamma_{SS}^\ell(-\mathbf{k})
= \pm [\Gamma_{SS}^\ell(-\mathbf{k})]^*
$
must be real for the intra-sublattice sectors and
$
\Gamma_{S \ne S'}^\ell(\mathbf{k}) = \pm \Gamma_{S' \ne S}^\ell(-\mathbf{k})
= \pm [\Gamma_{S \ne S'}^\ell(-\mathbf{k})]^*
$
for the inter-sublattice sectors, where the upper and lower signs 
correspond, respectively, to the singlet and triplet states. 
Considering the singlet states, the appropriate linearly-independent 
symmetry functions can be chosen as 
$
\Gamma_{AA}^a(\mathbf{k}) = 1 = \Gamma_{BB}^a(\mathbf{k})
$
with
$
C_{AA}^a = U = C_{BB}^a
$
for the intra-sublattice sectors, and
$
\Gamma_{AB}^a(\mathbf{k}) = \sqrt{2} \cos(k_x d) 
= \Gamma_{BA}^a(-\mathbf{k})
$
and
$
\Gamma_{AB}^b(\mathbf{k}) = \mathrm{i} \sqrt{2} \sin(k_x d) 
= \Gamma_{BA}^b(-\mathbf{k})
$
with
$
C_{AB}^a = V = C_{BA}^a 
$
and
$
C_{AB}^b = V = C_{BA}^b
$
for the inter-sublattice sectors. Similarly, considering 
the triplet states, the appropriate linearly-independent symmetry 
functions can be chosen as 
$
\Gamma_{AB}^a(\mathbf{k}) = \sqrt{2} \sin(k_x d) 
= -\Gamma_{BA}^a(-\mathbf{k})
$
and
$
\Gamma_{AB}^b(\mathbf{k}) = \mathrm{i} \sqrt{2} \cos(k_x d) 
= -\Gamma_{BA}^b(-\mathbf{k})
$
with
$
C_{AB}^a = V = C_{BA}^a 
$
and
$
C_{AB}^b = V = C_{BA}^b
$
for the inter-sublattice sectors. 

Equation~(\ref{eqn:LambdaSS}) is equivalent to a non-linear eigenvalue
problem for $E_{e \mathbf{q}}$. After recasting it as
$
\mathbf{G}_\mathbf{q} \boldsymbol{\Lambda}_{\mathbf{q}} = \mathbf{0},
$
we determine its self-consistent solutions by setting the eigenvalues 
of $\mathbf{G}_\mathbf{q}$ to zero one at a time. For instance, 
in the presence of two sublattices, i.e., $S = (A, B)$, and assuming
$\ell = (a,\{a,b\},a)$, respectively, for the $SS' = (AA, AB, BB)$ 
sectors as in the singlet case discussed above, the corresponding 
eigenvectors can be written as
$
\boldsymbol{\Lambda}_{\mathbf{q}} = (
\Lambda_{AA}^{a \mathbf{q}},
\Lambda_{AB}^{a \mathbf{q}}, \Lambda_{AB}^{b \mathbf{q}},
\Lambda_{BB}^{a \mathbf{q}}
)^\mathrm{T},
$
where $\mathrm{T}$ is the transpose. Note that, since the matrix 
elements that involve $\Lambda_{BA}^{\ell \mathbf{q}}$ are not 
independent, they are absorbed into the self-consistency equations 
via substitution by $\Lambda_{AB}^{\ell \mathbf{q}}$. As a result, 
for a given $\mathbf{q}$, we choose to label the resultant 
self-consistency solutions as $E_{e \mathbf{q}}$, where the label 
$e = \{1, 2, 3, 4\}$ indicates which eigenvalue of $\mathbf{G}_\mathbf{q}$
is set to $0$ starting with the lowest one. 
Similarly, assuming $\ell = \{a,b\}$ for the $AB$ sector of the 
triplet case discussed above, the corresponding eigenvectors 
can be written as
$
\boldsymbol{\Lambda}_{\mathbf{q}} = (
\Lambda_{AB}^{a \mathbf{q}}, 
\Lambda_{AB}^{b \mathbf{q}}
)^\mathrm{T},
$
leading to $E_{e \mathbf{q}}$ with $e = \{1, 2\}$. 
Thus, since the singlet (triplet) symmetry functions leads to a 
$4 \times 4$ ($2 \times 2$) nonlinear eigenvalue problem, 
Eq.~(\ref{eqn:LambdaSS}) gives rise to four (two) distinct singlet 
(triplet) branches. These six branches are shown in Fig.~\ref{fig:2body}
with different symbols. 

Our numerical benchmark shown in Fig.~\ref{fig:2body} clearly 
illustrates that bound-state solutions of Eq.~(\ref{eqn:alphanmk}) 
can be classified with respect to their exchange symmetry through 
the self-consistent solutions of Eq.~(\ref{eqn:LambdaSS}). 
Furthermore, it is pleasing to see that these results are in perfect 
agreement with the existing literature~\cite{nguenang09, valiente09}, 
with the caveat that the entire spectrum appears as folded into the 
BZ leading to the appearance of $4$ $(2)$ instead of $2$ $(1)$ 
singlet (triplet) branches. We also verified that the known analytical 
expression~\cite{nguenang09, kornilovitch24}
$
E_\mathbf{q}^\mathrm{triplet} = V + \frac{4t^2}{V} \cos^2(q_x d/2)
$
for the triplet branch in the usual BZ 
$
-\frac{\pi}{d} \le q_x \le \frac{\pi}{d}
$
is in perfect agreement with our numerical results. This expression is 
valid only when the energy of the triplet states are outside of the 
two-body continuum, i.e., it is not valid in the $V \to 0$ limit for 
which the triplet states are not allowed.

\section{Connection to quantum geometry}
\label{sec:geometry}

It is possible to relate our results to the recent literature on 
quantum-geometric effects in the formation of Cooper 
pairs~\cite{torma18, iskin21, iskin23, iskin24}. In the case of 
onsite interactions, this connection is known to be most transparent 
in a time-reversal symmetric system with a spatially-uniform order 
parameter in its unit cell. Motivated by this, we consider a system where
$
n_{S, -\mathbf{k}, \downarrow}^* = n_{S \mathbf{k} \uparrow} 
\equiv n_{S \mathbf{k}} 
$
and 
$
\varepsilon_{n, -\mathbf{k}, \downarrow} = 
\varepsilon_{n \mathbf{k} \uparrow} \equiv \varepsilon_{n \mathbf{k}}
$
are manifest. In addition, we assume that the pairing occurs primarily 
in one of the channels (say $\ell_0$th) with same coefficients
$
C_{S S'}^{\ell_0} = C_0^{\ell_0} < 0
$
for all of the nonzero interactions, and that the lowest-lying 
two-body bound states are described by the same amplitude
$
\Lambda_{S S'}^{\ell_0 \mathbf{q}} = \Lambda_0^{\ell_0 \mathbf{q}}
$ 
in the small-$\mathbf{q}$ regime. We note that the same construction 
applies to the highest-lying bound states when $C_0^{\ell_0} > 0$.
Under these assumptions, Eq.~(\ref{eqn:LambdaSS}) reduces to 
\begin{align}
1 = 
-\frac{C_0^{\ell_0}}{N_c N_0} 
\sum_{nm\mathbf{k}}
\frac{
\langle m_{\mathbf{k}-\frac{\mathbf{q}}{2}}
| (\mathcal{L}_\mathbf{k}^{\ell_0})^\mathrm{T}|
n_{\mathbf{k}+\frac{\mathbf{q}}{2}}\rangle
\langle n_{\mathbf{k}+\frac{\mathbf{q}}{2}}
| (\mathcal{L}_\mathbf{k}^{\ell_0})^*|
m_{\mathbf{k}-\frac{\mathbf{q}}{2}}\rangle
}
{\varepsilon_{n, \mathbf{k} +\frac{\mathbf{q}}{2}}
+ \varepsilon_{m, \mathbf{k}-\frac{\mathbf{q}}{2}} 
- E_{0 \mathbf{q}}},
\label{eqn:E0q}
\end{align}
where $N_0 = \sum_{SS'}^{'} 1$ is the number of nonzero 
$\Lambda_{S S'}^{\ell_0 \mathbf{q}}$ parameters in the system, 
$| n_\mathbf{k} \rangle$ is the state vector in the sublattice 
basis, and $E_{0 \mathbf{q}}$ is the energy of the lowest-lying 
bound state. The nonzero matrix elements of 
$\mathcal{L}_\mathbf{k}^{\ell_0}$ are the symmetry factors 
$\Gamma_{SS'}^{\ell_0}(\mathbf{k})$ of those sublattice sectors 
whose $\Lambda_{S S'}^{\ell_0 \mathbf{q}}$ are nonzero, and 
$\mathrm{T}$ is the transpose. Equation~(\ref{eqn:E0q}) is
the generalization of our previous result under the so-called
uniform-pairing condition~\cite{iskin23, iskin24}.

Within this construction, in the case when there is an 
energetically-isolated flat band with energy $\varepsilon_f$ in 
the Bloch spectrum, its low-energy bound states simply follow from
Eq.~(\ref{eqn:E0q}), leading to
$
E_{0 \mathbf{q}} = 2\varepsilon_f + 
\frac{C_0^{\ell_0}}{N_c N_0} \sum_\mathbf{k} 
\langle f_{\mathbf{k}-\frac{\mathbf{q}}{2}}
| (\mathcal{L}_\mathbf{k}^{\ell_0})^\mathrm{T}|
f_{\mathbf{k}+\frac{\mathbf{q}}{2}}\rangle
\langle f_{\mathbf{k}+\frac{\mathbf{q}}{2}}
| (\mathcal{L}_\mathbf{k}^{\ell_0})^*|
f_{\mathbf{k}-\frac{\mathbf{q}}{2}}\rangle
$
in the small-$\mathbf{q}$ regime.
For instance, in the particular case when $\mathcal{L}_\mathbf{k}^{\ell_0}$ 
is a diagonal matrix with isotropic elements in all sublattice sectors, 
i.e.,
$
\Gamma_{SS'}^{\ell_0}(\mathbf{k}) = \Gamma_0^{\ell_0}(\mathbf{k}) \delta_{SS'},
$
we find 
\begin{align}
E_{0 \mathbf{q}} = 2\varepsilon_f + 
\frac{C_0^{\ell_0}}{N_c N_b} \sum_\mathbf{k} 
[\Gamma_0^{\ell_0}(\mathbf{k})]^2
\big |\langle f_{\mathbf{k}-\frac{\mathbf{q}}{2}}
| f_{\mathbf{k}+\frac{\mathbf{q}}{2}}\rangle \big|^2.
\end{align}
According to Eq.~(\ref{eqn:beta}), this case corresponds to a multiband 
lattice whose intra-sublattice order parameter 
$
\beta_{0 \mathbf{k}}^\mathbf{q} = 
\Lambda_0^{\ell_0 \mathbf{q}} \Gamma_0^{\ell_0}(\mathbf{k})
$
is the same for all sublattices in the small-$\mathbf{q}$ regime. 
Then, the geometric contribution to the effective-band mass of the 
lowest-lying bound states becomes apparent upon the use of 
power-series expansion
$
|\langle n_\mathbf{k} | m_\mathbf{k-q} \rangle|^2 = \delta_{nm} - 
\frac{1}{2} \sum_{ij} 
\big[g_{ij}^{n\mathbf{k}} \delta_{nm} 
+ g_{ij}^{nm\mathbf{k}}(\delta_{nm} - 1) \big] q_i q_j
$
in $\mathbf{q}$, where 
$
g_{ij}^{nm\mathbf{k}} = 2\mathrm{Re} 
\langle \dot{n}_\mathbf{k}^i | m_\mathbf{k} \rangle
\langle m_\mathbf{k} | \dot{n}_\mathbf{k}^j \rangle
$
is the band-resolved quantum metric and
$
g_{ij}^{n\mathbf{k}} = \sum_{m \ne n} g_{ij}^{nm\mathbf{k}}
$
is the quantum metric of the $n$th band~\cite{iskin24}.
Here $\mathrm{Re}$ denotes the real part, 
$
| \dot{n}_\mathbf{k}^i \rangle = \partial | n_\mathbf{k} \rangle/\partial k_i,
$ 
and $\delta_{ij}$ is a Kronecker delta. Thus, we recover the well-known 
result for the onsite interactions when $C_0^{\ell_0} = U$ and 
$\Gamma_0^{\ell_0}(\mathbf{k}) = 1$~\cite{torma18}. In addition, 
we numerically verified that our analysis above applies perfectly well 
to the Creutz lattice with nearest-neighbor interactions where 
$C_0^{\ell_0} = V$ and
$
\Gamma_0^{\ell_0}(\mathbf{k}) = \sqrt{2} \cos (k_x d)
$ 
for the intra-sublattice sectors. Finally, we remark in passing 
that, if the flat band is not isolated or in the general case when 
there are dispersive bands, it is possible to derive the quantum-geometric
contribution to the effective-band mass, within our construction, 
by directly expanding Eq.~(\ref{eqn:E0q}) in powers of 
$\mathbf{q}$~\cite{iskin23, iskin24}.

\section{Conclusion}
\label{sec:conc}

In summary, here we analyzed the two-body problem within a generic multiband 
extended-Hubbard model, including finite-ranged hopping and interaction
parameters. In particular, we derived self-consistency relations 
for the two-body bound states using an exact variational approach, which can 
be easily applied to various lattice geometries. To validate their accuracy 
numerically, we compared our results to the existing literature on the 
linear-chain model. Our findings demonstrated perfect agreement between 
the spin singlet and triplet states obtained through our method and those 
reported in the literature. As an outlook, it would be intriguing to apply 
the recently proposed bulk-edge correspondence for the nonlinear eigenvalue 
problems to the two-body bound states by introducing their auxiliary 
eigenvalues~\cite{isobe24}. Furthermore, one can also study the Chern 
numbers for the triplet bound states by following our recent work on singlet
bound states for the onsite Hubbard model~\cite{alyuruk24}, i.e., 
by utilizing the eigenvectors $\boldsymbol{\Lambda}_{\mathbf{q}}$ of the 
nonlinear eigenvalue problem. Finally, in the spinless case, the two-body 
bound states for the extended-Hubbard and extended-Bose-Hubbard models 
can be studied through our triplet and singlet solutions, respectively,
by suppressing the spin labels.

\begin{acknowledgments}
The author acknowledges funding from US Air Force Office of Scientific 
Research (AFOSR) Grant No. FA8655-24-1-7391.
\end{acknowledgments}

\bibliography{refs}

\end{document}